\documentclass[a4paper,pra,twocolumn]{revtex4}
\usepackage{graphicx}
\usepackage{epsfig,amsmath,amssymb,mathtools,amscd}
\usepackage[linkcolor=red]{hyperref} 
\usepackage{dsfont}
\usepackage{subfigure}
\usepackage{enumerate}
\newcommand{\Tr}{\mathop{\mathrm{Tr}}}
\def\ie{{i.~e.}}
\mathchardef\minus="002D

\def\<{\langle}\def\>{\rangle}

\def\Reals{\mathbb{R}}

\def\transf#1{\mathcal{#1}}
\def\tA{\transf A}
\def\tI{\transf I}
\def\L2{{\mathcal L}_2}

\def\d#1 {\mathop{\!\! \mathrm{d}#1}\,}
\def\df#1#2 {\!\!\frac{\mathop{\mathrm{d}#1}}{#2}\,}
\def\SU{\mathbb{SU}}

\def\bvec#1{\mathbf{#1}}
\def\bk{\bvec k}
\def\bh{\bvec h}

\def\bn{\bvec n}

\def\bg{\bvec g}

\def\bn{\bvec n}

\def\d{\operatorname{d}}
\def\Exp{\operatorname{Exp}}

\begin{document}
\title{Quantum cellular automata and free quantum field theory}

\author{Giacomo Mauro D'Ariano}
\author{Paolo Perinotti}



\date\today

\begin{abstract} 
In a series of recent papers \cite{PhysRevA.90.062106,bisio2013dirac,Bisio2015244,bisio2014quantum} 
it has been shown how free quantum field theory can be derived without using mechanical primitives (including space-time, special relativity, quantization rules, etc.), but only considering the easiest quantum algorithm encompassing a countable set of quantum systems whose network of interactions satisfies the simple principles of unitarity, homogeneity, locality, and isotropy. This has opened the route to extending the axiomatic information-theoretic derivation of the quantum theory of abstract systems \cite{quit-derivation} to include quantum field theory. The inherent discrete nature of the informational axiomatization leads to an extension of quantum field theory to a quantum cellular automata theory, where the usual field theory is recovered in a regime where the discrete structure of the automata cannot be probed. A simple heuristic argument sets the scale of discreteness to the Planck scale, and the customary physical regime where discreteness is not visible is the relativistic one of small wavevectors. 

In this paper we provide a thorough derivation from principles that in the most general case the graph of the quantum cellular automaton is the Cayley graph of a finitely presented group, and showing how for the case corresponding to  Euclidean emergent space (where the group resorts to an Abelian one) the automata leads to Weyl, Dirac and Maxwell field dynamics in the relativistic limit. We conclude with some perspectives towards the more general scenario of non-linear automata for interacting quantum field theory.
\end{abstract}

\maketitle

\section{Introduction}

Since its very beginning, quantum information theory has represented a new way of looking at foundations of Quantum Theory (QT), and the study of quantum protocols has provided a significant reconsideration of the of the structure of the theory, which eventually resulted in a new axiomatization program, initiated in the early 2000 
\cite{hardy2001quantum,fuchs2002quantum,dariano114,d2010probabilistic}. The purpose was to reconstruct the  von Neumann Hilbert-space formulation of the theory starting from information-processing principles. A complete derivation of QT for finite dimensions has been finally achieved in Ref. \cite{quit-derivation} within the framework of operational probabilistic theories, starting from six principles assessing the possibility or impossibility to carry out specific information-processing tasks.

As a theory of information processing, however, QT does not carry any physical semantics or mechanical notions--such as space-time, elementary particles, mass, charge--nor physical constants as the Planck constant and the speed of light. The program now aims at recovering also the mechanical features, instead of following the historical approach of imposing quantization rules and mysteriously turning classical Hamiltonians to quantum. The informational approach is pursued even further, with the purpose of reconstructing also the quantum equations of motion, which in the simplest non interacting case are the Weyl, Dirac and Maxwell field theories, along with recovering the fundamental constants, such as $\hbar$ and $c$. The starting idea is to look at physical laws as an effective description of an information processing algorithm, 
which updates the states of an array of quantum memory cells, with particles emerging as the interpretation of special patterns of the memory. It is important to stress that space-time itself is also emergent in this approach, as the natural set of coordinates in which the emergent dynamics is formulated. 

The present follow-up of the informational derivation of QT is also motivated by the the role that information is playing in theoretical physics at the fundamental level of quantum gravity and Planck-scale, e.~g. in the holographic-principle and the ultraviolet cutoffs, implying an upper bound to the amount of information that can be stored in a finite space volume
\cite{bekenstein1973black,hawking1975particle,PhysRevLett.90.121302}. Imposing an {\em in-principle} upper bound to the information density, forces us to replace continuous quantum fields with countably many finite-dimensional quantum systems, i.e. a quantum cellular automaton (QCA)~\cite{feynman1982simulating} representing the unitary
evolution of the quantum systems in local interaction \cite{grossing1988quantum,aharonov1993quantum,ambainis2001one}. 
 
The possibility of approximating relativistic quantum dynamics with QCAs was already known from 
Refs.~\cite{bialynicki1994weyl,meyer1996quantum,Yepez:2006p4406} However, the possibility of 
reversing the paradigm and deriving the equations of quantum field theory (QFT) from informational principles was proposed by one of the present authors only in recent years in a series of heuristic works  
\cite{darianopirsa,darianovaxjo2010,darianovaxjo2011,darianovaxjo2012,mauro2012quantum,dariano-asl,darianosaggiatore}, which preluded the main work \cite{PhysRevA.90.062106} from the present authors with the derivation Weyl and Dirac, along with the works \cite{bisio2013dirac,Bisio2015244} and the derivation of Maxwell \cite{bisio2014quantum}. Other authors then also addressed QFT in the QCA framework~\cite{arrighi2014dirac,farrelly2013discrete}.

The QCA framework manifestly breaks the Lorentz covariance, and the claim that relativistic quantum field theory is recovered must be substantiated by an appropriate analysis of the symmetries of the emerging space-time. To this end, one has to introduce the notion of ``inertial frame'' in terms of the underlying QCA without using space-time. Upon identifying the notion of ``reference frame'' with that of ``representation'' of the dynamics, we appeal to the relativity principle to define the ``inertial representation'' as the one for which the physical law retains the same mathematical form. In such a way the change of inertial reference frame leads to a set of modified Lorentz transformations that recover the usual ones when the observation scale is much larger than the discrete microscopic scale. This problem has been first addressed in Refs.~\cite{bibeau2013doubly,lrntz3d}. While the QCA model recovers the usual Poincar\'e covariance of QFT in the relativistic limit of wave-vectors much smaller than Planck's one\cite{Bisio2015244,PhysRevA.90.062106,bisio2014quantum} (namely in the limit where discreteness cannot be probed), the group of symmetries exhibits a very different behavior in the ultra-relativistic regime of Planckian wave-vectors, where the usual symmetries are distorted as in doubly-special relativity models.

In this paper we review the derivation from principles of
Refs.~\cite{PhysRevA.90.062106,bisio2014quantum}, proving in detail that the graph of the QCA is a   \emph{Cayley graph} of a finitely presented group, and showing how for the case corresponding to an Euclidean emergent space (where the group resorts to an Abelian one) the automata lead to Weyl, Dirac and Maxwell field dynamics in the relativistic limit. We conclude with some perspectives towards the more general scenario of non-linear automata for interacting quantum field theory. 

\section{The principles for the QCA}\label{s:assumptions}

A QCA gives the evolution of a denumerable set $G$ of cells, each one corresponding to a quantum system.  In our framework (see Refs.~\cite{Bisio2015244,PhysRevA.90.062106}) we are interested in exploring the possibility of an automaton description of free QFT and thus assume the quantum systems in $G$ to be quantum fields. Requiring that the amount of information in a finite number of cells must be finite corresponds to consider \emph{Fermionic} modes. In Section \ref{s:maxwell}, based on
Ref.~\cite{bisio2014quantum}, we see how \emph{Bosonic} statistics can be recovered in this scenario as a very good approximation with the bosonic mode corresponding to a specially entangled state of a pair of Fermionic modes. The relation between Fermionic modes and finite-dimensional quantum systems, say \emph{qubits}, is studied in the literature, and the two theories were proved to be computationally equivalent \cite{Bravyi2002210}. On the other hand the quantum theory of qubits and the quantum theory of Fermions differ in the notion of what are local transformations \cite{doi:10.1142/S0217751X14300257,0295-5075-107-2-20009},
with local Fermionic operations mapped into nonlocal qubit transformatioms and vice versa. 

From now each cell of $G$ will host an array of Fermionic modes with field
operator $\psi_{g,l}$, obeying the canonical anti-commutation relations
\begin{equation}
\{\psi_{g,l},\psi_{g',l'}\}=0,\quad\{\psi_{g,l},\psi^\dag_{g',l'}\}=\delta_{g,g'}\delta_{l,l'},
\end{equation}
where $l=1,\ldots, s_g$, $s_g$ denotes the number of field components
of the array $\psi_g$ at each site $g\in G$. The general states and effects are linear combinations of even products of field operators (see Ref.\cite{doi:10.1142/S0217751X14300257}). The evolution occurs in discrete identical steps, and in each one every cell interacts with the others. The construction of the one-step update rule is based on the following assumptions on the interactions among systems\cite{PhysRevA.90.062106} : 1) unitarity, 2) linearity, 3) homogeneity, 4) locality, and 5) isotropy. 
These constraints regard the algebraic properties of the map providing the update rule of the field. Denoting the variable that counts the evolution steps by $t$, and the local array of field operators at $g$ at step $t$ by $\psi_{g,t}$ we can express unitarity as follows
\begin{align}
\psi_{g,t+1}=\tA\psi_{g,t}:=U\psi_{g,t}U^\dag,
\end{align}
with $U$ unitary operator. The linearity constraint requires that the field evolution can be expressed in terms of  linear combinations of field operators, namely
\begin{align}
\psi_{g,t+1}=\sum_{g'}A_{g,g'}\psi_{g',t},
\end{align}
where $A_{g,g'}$ is an $s_g\times s_{g'}$ complex matrix called {\em transition matrix}. Linearity thus endows the set $G$ with a graph structure $\Gamma(G,E)$, with vertex set $G$ and edge set  $E=\{(g,g')|A_{g,g'}\neq0\}$. For every $g\in G$, we define the set $S_g:=\{A_{g,g'}\neq0\}$ of non-null transition matrices, along with the {\em neighborhood} of $g$ as $N_g:=\{g'\in G|A_{g,g'}\neq 0\}$.

Homogeneity consists in the requirement that every two vertices are indistinguishable. The most general  discrimination procedure between two vertices occurs in a finite number $N$ of steps and consists of a suitable sequence of state preparations of local modes, at different steps, followed by a sequence of measurements. A necessary condition for homogeneity is thus the following: for every vertex $g\in G$ the array $\psi_g$ has the same length, $s_g=s$. If we now consider a general permutation $\pi$ of the vertices, we will denote by $w_\pi$ the transformation defined by $w_\pi(\psi_{g,l})=\psi_{\pi(g),l}$. Homogeneity can thus be expressed as the requirement that for every $g,g'$ there exists $\pi$ such that $\pi(g)=g'$, and for every joint state $\rho$ and every joint effect $O$ of the automaton along with a generic ancillary system ${\rm R}$, one has
\begin{align}
&\Tr[\rho (\tA\otimes \tI_{\rm R}) (O)]\nonumber\\
&\quad=\Tr[(w_\pi^\vee\otimes\tI_{\rm R})(\rho) (\tA\otimes\tI_{\rm R}) \{(w_\pi\otimes\tI_{\rm R})(O)\}].
\label{eq:homogen}
\end{align}
where $\tI_{\rm R}$ denotes the identical transformation on the ancillary system ${\rm R}$.
As one can easily verify, the dual map $w_{\pi}^\vee$ coincides with $w_{\pi^{-1}}$.
The first result that we show is thus the following equivalent condition for homogeneity: 
a cellular automaton $\tA$ on the set $G$ is homogeneous if and only if for every $g,g'\in G$ there exists a permutation 
$\pi:G\to G$ such that $\pi(g)=g'$, and
\begin{align}
w_{\pi^{-1}}\tA w_\pi=\tA.
\label{eq:commut}
\end{align}
It is easy to check that Eq.~\eqref{eq:commut} equally holds for $\tA^N$, for any $N>0$. 
The permutations $\pi$ that satisfy condition \eqref{eq:commut} are clearly a group $\Pi$ that acts transitively on $G$.

%

Considering a general element $\psi_{g}$, the condition in Eq.~\eqref{eq:commut} implies that for some $\pi\in\Pi$
\begin{align}
&\sum_{f'\in N_{\pi(g)}}A_{\pi(g)f'}\psi_{\pi^{-1}(f')}=\sum_{f\in N_{g}}A_{gf}\psi_{f}.
\label{eq:perms}
\end{align}
Since the field operators are linearly independent, Eq.~\eqref{eq:perms} bears two important consequences: for every $f\in N_{g}$ there exists $f'\in N_{\pi(g)}$ such that $\pi^{-1}(f')=f$---or equivalently $f'=\pi(f)$---and viceversa for every $f'\in N_{\pi(g)}$ there exists $f\in N_{g}$ such that $f'=\pi(f)$. Thus, $N_{\pi(g)}=\pi(N_{g})$, and since the group of permutations $\pi$ satisfying Eq.~\eqref{eq:commut} is transitive, we have that for every $g,g'$ there is a bijection $N_g\leftrightarrow N_{g'}$. Setting $N:=N_{\bar g}$, for every $g\in G$ one has a bijection $N_g\leftrightarrow N$.

Moreover, by Eq.~\eqref{eq:perms}, for every $g$ and for every $f\in N_{g}$, one has $A_{gf}= A_{\pi(g)\pi(f)}$ for the permutations $\pi$ satisfying Eq.~\eqref{eq:commut}. Again, since the group of such permutations is transitive on $G$, for every pair $g,g'\in G$ the sets $S_g$ and $S_{g'}$ contain the same $s\times s$ transition matrices, namely $S:=S_g=S_{g'}=\{A_{h_1}\}_{i=1}^{|N|}$. If we associate the label $h_i$ to the edge $(g,g')$ whenever $A_{g,g'}=A_{h_i}$, we enrich the structure of the graph $\Gamma(G,E)$, which becomes a vertex-transitive colored directed graph, with colors corresponding to the labels $h_i$. If two transition matrices $A_{h_1}=A_{h_2}$ are equal, we conventionally associate them with two different labels $h_1\neq h_2$ in such a way that Eq.~\eqref{eq:perms} holds. If such choice is not unique, we will pick an arbitrary one, since the homogeneity requirement implies that there exists a choice of labeling for which all the following construction is consistent. In the following we will identify the set $S$ with the set of labels $h_i$, with a slight abuse of notation. We now define the action of $S$ on $G$ formally as $g'=gh_i$ when $A_{gg'}=A_{h_i}$. Notice that by construction, one has $A_{\pi(g)\pi(f)}=A_{gf}=A_{h_i}$, which implies
\begin{align}
\pi(g)h_i=\pi(f)=\pi(gh_i).
\label{eq:action}
\end{align}


If we now use the alphabet $S\cup S^{-1}$ of labels $h_i$ and $h_{i}^{-1}$ to form arbitrary words, we obtain a free group $F$: composition corresponds to word juxtaposition, with the empty word $\lambda$ representing the identity, and the formal rule $h_ih_i^{-1}=h_i^{-1}h_i=\lambda$. An element $w=h_{i_1}^{p_1}h_{i_2}^{p_2}\ldots h_{i_n}^{p_n}$ of $F$---with $p_j\in\{-1,1\}$---thus corresponds to a path on the graph, where the symbol $h_i^{-1}$ denotes a backwards step along an arrow (i.e.~from the head of the arrow to its tail). For every $h_{i_1}^{p_1}h_{i_2}^{p_2}\ldots h_{i_m}^{p_m}=w\in F$, one has $w^{-1}=h_{i_m}^{-p_m}\ldots h_{i_2}^{-p_2}h_{i_1}^{-p_1}$. The action of symbols $h_i\in S$ on the elements $g\in G$ can now be extended to arbitrary words $w\in F$, by posing $gh_{i}^{-1}=g'$ iff $g'h_i=g$, and $g h_{i_1}^{p_1}h_{i_2}^{p_2}:=(g h_{i_1}^{p_1})h_{i_2}^{p_2}$. For every $w\in F$, and for every pair $g,g'\in G$ (for the corresponding permutation $\pi$), we now show that $\pi(fw)=\pi(f)w$. The first step consists in proving the result for $w=h_i^{-1}$. Let $f'=fh_i^{-1}$, namely $f=f'h_i$. Then by Eq.~\eqref{eq:action} $\pi(f)=\pi(f')h_i$, namely $\pi(f)h_i^{-1}=\pi(f')=\pi(fh_i^{-1})$. Notice that, if we define $N'_g:=\{g'|g\in N_{g'}\}$, the last result implies that for every pair $f,g\in G$ there is a bijection $N'_{f}\leftrightarrow N'_g$, and $N'_{\pi(g)}=\pi(N'_g)$. Indeed, $g'\in N'_{g}$ if and only if $g'=gh^{-1}_j$ for some $j$, and thus $\pi(g')=\pi(g)h_j^{-1}\in N'_{\pi(g)}$. One can prove that $\pi(fw)=\pi(f)w$ by induction on the length $l(w)$ of the word $w$. Indeed, we know that it is true for $l(w)=1$. Suppose now that for $l(w)=n-1$ one has $\pi(fw)=\pi(f)w$, and consider $w'$ with $l(w')=n$. Then $w'=wh^{p}_i$ with $l(w')=n-1$ and $p=\pm1$. In this case we have 
\begin{align*}
\pi(fw')&=\pi(fwh_{i}^{p})\\
&=\pi[(fw)h_{i}^{p}]\\
&=\pi(fw)h_{i}^{p}\\
&=\pi(f)wh_{i}^{p}\\
&=\pi(f)w',
\end{align*}
where the induction hypothesis is used in the fourth equality.

Let us now suppose that for some $f\in G$ and some word $w\in F$ one has $fw=f$. Then for every $f'\in G$ one can take $\pi$ such that $\pi(f)=f'$, thus obtaining
\begin{align}
f'w=\pi(f)w=\pi(fw)=\pi(f)=f'.
\label{eq:clospath}
\end{align}
Thus, if a path $w\in F$ is closed starting from $f\in G$, then it is closed also starting from any other $g\in G$.

In particular, the necessary condition implies that if for some $h_i\in S$, there exists an element $h_j\in S$ and $g,g'\in G$ such that $A_{gg'}=A_{h_i}$ and $A_{g'g}=A_{h_j}$, then for every $f\in G$ one has $fh_ih_j=fh_jh_i=f$, namely $h_j=h_i^{-1}$. We can now easily see that the subset $R$ of $F$ corresponding to words $r$ such that $gr=g$ for all $g\in G$ is a normal subgroup. Indeed, $R$ is a subgroup because the juxtaposition of two words $w,w'\in R$ is again  a word $ww'\in R$, and for every word $w\in R$ also $w^{-1}\in R$. To prove that $R$ is normal in $F$ we just show that it coincides with its normal closure, i.e.~for every $w\in F$ and every $r\in R$, we have $wrw^{-1}\in R$. Indeed, defining for arbitrary $g$ the element $g':=gw$, we have $g'w^{-1}=g$, and thus $gwrw^{-1}=g'rw^{-1}=g'w^{-1}=g$, namely $wrw^{-1}\in R$. 

We thus identified a normal subgroup $R$ containing all the words $r$ corresponding to closed paths. If one takes the quotient $F/R$, one obtains a group whose elements are equivalence classes of words in $F$. If we label an arbitrary element of $G$ by $e$, it is clear that the elements of $G$ are in one-to-one correspondence with the vertices of $G$, since for every $g\in G$ there is one and only one class in $F/R$ whose elements lead from $e$ to $g$. We can then write $g=w$ for every $w\in F$ such that $w$ represents a path leading from $e$ to $g$. In technical terms, the graph $\Gamma(G,E)=\Gamma(G,S)$ is the Cayley graph of the group $G=F/R$. Homogeneity thus implies that the set $G$ is a group $G$ that can be presented as $G=\<S|R\>$, where $S$ is the set of {\em generators} of $G$ and $R$ is the group of {\em relators}. In the following, if $h_i=h_i^{-1}$ we will draw an undirected edge to represent $h_i$. The presentation can be chosen by arbitrarily dividing $S$ into $S_+\subseteq S$ and $S_-:=S_+^{-1}$ in such a way that $S_+\cup S_-=S$. The above arbitrariness is inherent the very notion of group presentation and corresponding Cayley graph, and will be exploited in the following, in particular in the definition of isotropy.

For convenience of the reader we remind the definition of \emph{Cayley
  graph}. Given a group $G$ and a set $S_+$ of generators of the group,
the Cayley graph $\Gamma(G,S_+)$ is defined as the colored directed
graph having vertex set $G$, edge set $\{(g,gh);g\in G, h\in S_+\}$, and
a color assigned to each generator $h\in S_+$. Notice that a Cayley
graph is \emph{regular}---i.e.~each vertex has the same degree---and
\emph{vertex-transitive}---i.e.~all sites are equivalent, in the sense that
the graph automorphism group acts transitively upon its
vertices. The Cayley graphs of a group $G$ are in one to one
correspondence with its presentations, with $\Gamma(G,S_+)$
corresponding to the presentation $\<S_+|R\>$. We finally
remind that a Cayley graph is said to be \emph{arc-transitive} when its
group of automorphisms acts transitively not only on its vertices but
also on its directed edges.

\begin{figure}[ht]
\includegraphics[width=6cm]{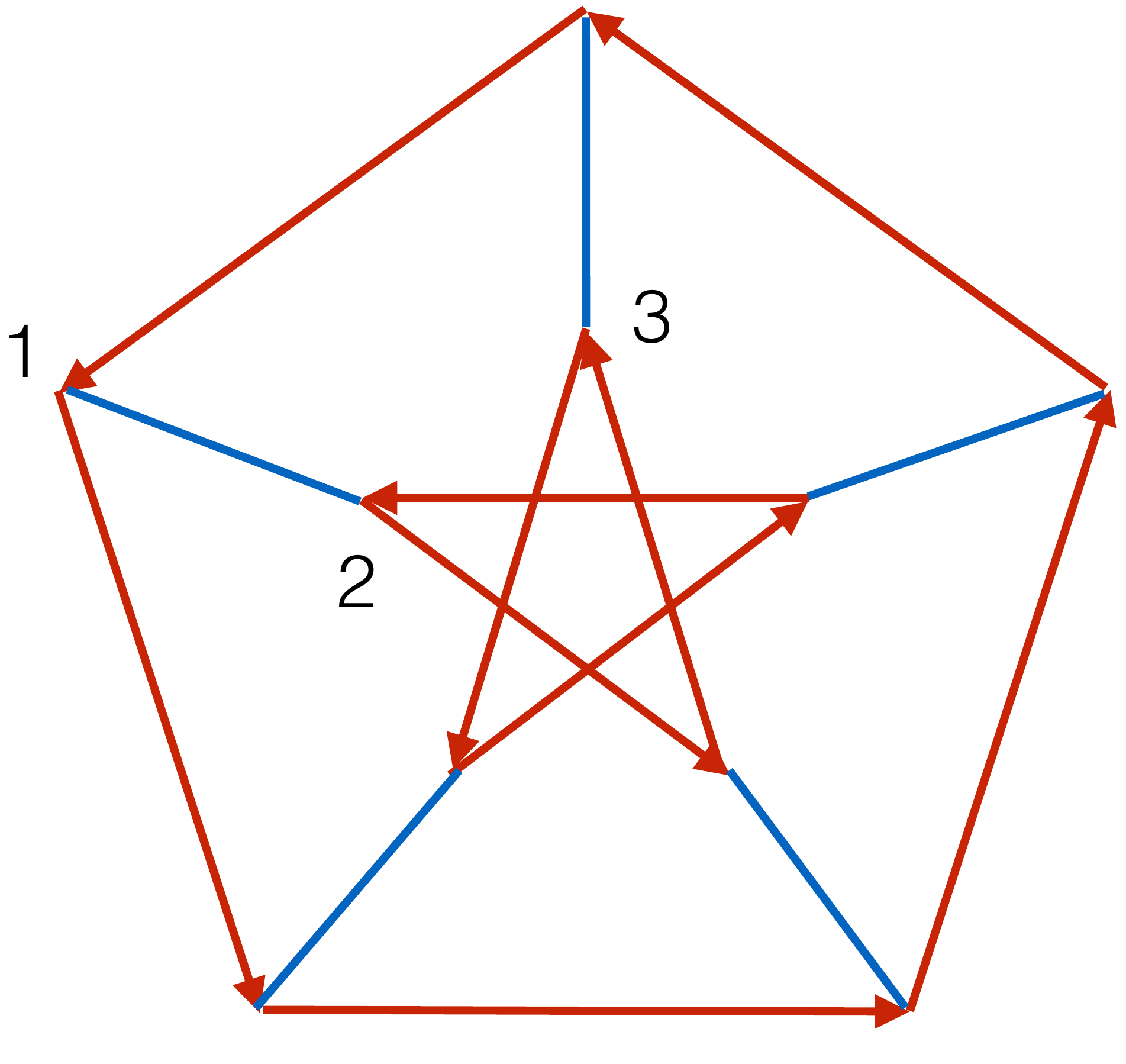}
\caption{x}
\label{f:peters}
\end{figure}

Notice that the sole property of vertex transitivity, without the necessary condition that closed paths are the same starting from every vertex [i.e.~Eq.~\eqref{eq:clospath}], would not be sufficient to identify a group structure. Consider indeed the Petersen graph in Fig.~\ref{f:peters}, whose vertices are equivalent. It is known that the Petersen graph cannot represent a Cayley graph, and this is due to the failure of the condition on closed paths. One can easily verify that, up to irrelevant permutations, the Petersen graph can be directed and colored in a unique way, that is the one in Fig.~\ref{f:peters}. Now, the path $brrbr$ is closed starting from vertex 1, while it leads from vertex 2 to vertex 3.

We can now easily prove that if a linear cellular automaton $\tA$ has the property that its transition matrices are independent of the system $g$, \ie~$S_g=S=\{A_{h_i}\}_{i=1}^{|N|}$, and they define the Cayley graph of a group, then $\tA$ is homogeneous. Indeed, in this case one can define for $g,g'$ the permutation $\pi(f):=g'g^{-1}f$, which clearly gives $\pi(g)=g'$, with $\pi^{-1}(f)=gg'^{-1}f$. In this case, one has
\begin{align*}
w_{\pi^{-1}}\tA w_\pi(\psi_g)&=w_{\pi^{-1}}\tA(\psi_{\pi(g)})\\
&=w_{\pi^{-1}}\sum_{f\in N_{\pi(g)}}A_{\pi(g)f}\psi_f\\
&=w_{\pi^{-1}}\sum_{f\in N_{g}}A_{\pi(g)\pi(f)}\psi_{\pi(f)}\\
&=w_{\pi^{-1}}\sum_{h\in S}A_{h}\psi_{\pi(g)h}\\
&=\sum_{h\in S}A_{h}\psi_{\pi^{-1}(g'h)}\\
&=\sum_{h\in S}A_{h}\psi_{\pi^{-1}(g')h}\\
&=\sum_{h\in S}A_{h}\psi_{gh}\\
&=\tA (\psi_g),
\end{align*}
which implies the homogeneity condition of Eq.~\eqref{eq:homogen}.

Locality is the requirement that the cellular automaton can be determined by preparing and measuring a finite number of systems after they evolve for a finite number of steps. Notice that determining a homogeneous cellular automaton amounts to determine the set $S$ of transition matrices along with the set $R$ of closed paths, which characterizes the group $G$. If $S$ has to be determined by measurements on a finite number of systems, then the set $S$ has to be finite. For a similar reason, the set $R$ must be completely determined by a finite set of closed paths of finite size. This implies that the group $G$ must be finitely presented. In terms of the evolution rule, every local Fermionic system interacts with a finite number of other systems at each step.

The automaton can then be represented by an operator over the Hilbert space $\ell^2(G)\otimes \mathbb C^s$
\begin{equation}\label{eq:noniso}
  A=\sum_{h\in S} T_h\otimes A_h,
\end{equation}
where $T$ is the {\em right-regular} representation of $G$ on $\ell^2(G)$,
$T_g|g'\>=|g'g^{-1}\>$.

We remind now that the set $S$ can be split in many ways as $S=S_+\cup S_-$, with $\{e\}$ denoting the identity in $G$, that appears only in the presence of self-interaction. The requirement of \emph{isotropy} amounts to the statement that all directions on $\Gamma(G,S_+)$ are equivalent. This requirement is translated in mathematical terms requiring that there exists a decomposition of $S=S_+\cup S_-$, and a faithful representation $U$ over $\mathbb{C}^s$ of a group $L$ of graph automorphisms that is transitive over $S_+$, such that
one has the covariance condition
\begin{equation}\label{eq:covW}
A=\sum_{h\in S} T_h\otimes A_h=\sum_{h\in S} T_{l(h)}\otimes U_lA_hU^\dag_l,\quad \forall l\in L.
\end{equation}
By linear independence of the generators $T_h$ of the right regular representation of $G$ one has that the above condition \ref{eq:covW} implies
\begin{align}
A_{l(h^{\pm1})}=U_l A_{h^\pm1} U_l^\dag.
\end{align}
Notice that, as a consequence of this assumption, the Cayley graph $\Gamma(G,S_+)$ must be arc-transitive. Notice also that the same automaton on the Cayley graph corresponding to the presentation $G=\<S|R\>$ might in principle satisfy isotropy for one or more choices of the set $S_+$ and group $L$. For a given $S$, different choices of $S_+$ correspond to different orientations of some edges over the same colored graph. However, in the special cases that we consider here, the choice of representation satisfying the isotropy requirement turns out to be unique.

A covariant automaton of the form \eqref{eq:covW} describes the free
evolution of a field by a quantum algorithm with finite algorithmic
complexity, and with homogeneity and isotropy corresponding to the universality of the law given by the algorithm. 

As a consequence of the assumptions, the \emph{unitarity} condition---imposing that the map $\tA$ is unitary---is given by  
\begin{align}
  &\sum_{h\in S}A^\dag_h A_h=\sum_{h\in S}A_h A^\dag_h=I_s,\nonumber\\
  &\sum_{\shortstack{$\scriptstyle h,h'\in S$\\ $\scriptstyle
      h^{-1}h'=h''$}} A^\dag_h A_{h'}=\sum_{\shortstack{$\scriptstyle
      h,h'\in S$\\ $\scriptstyle h'h^{-1}=h''$}} A_{h'} A^\dag_{h}=0
      \label{eq:unitarity}
\end{align}
in terms of the transition matrices $A_h$.

\section{Emergent spacetime}\label{s:spacetime}

In the previous Section we have seen how our assumptions lead to a
model of evolution on a discrete computational space endowed with the
structure of Cayley graph. The usual dynamics on continuous spacetime
is expected to emerge as an effective description that holds in the
regimes where the discrete scale cannot be probed.  

Within this perspective space and time emerge from the structure of the graph 
with the time variable corresponding to the computational step of the automaton. 
The automaton represents a physical law, giving rise to a picture of phenomena in 
a spacetime $M$ within a given reference frame corresponding to the description of a specific observer.
The spacetime $M$ has a Cartesian product structure $M=X\times T$, with $T$ the one
dimensional manifold corresponding to time (clearly diffeomorphic to
the real line) and the $X$ the (generally $n$-dimensional) manifold
representing space. The spacetime manifold $M$ here
emerges as described in a fixed reference frame. The notion of change of
reference frame based on the invariance of the QCA dynamics was studied in 
Refs.~\cite{bibeau2013doubly,lrntz3d}. The steps
of the automaton evolution can be represented as a totally ordered set
of points $t_1,t_2,\ldots$ with the metric
$d_t(t_i,t_j)=|j-i|$. Similarly on the graph we take the metric $d_x$
induced by the word-counting on the Cayley graph.

\begin{figure}[h!]
  \begin{center}
    \includegraphics[width=8cm]{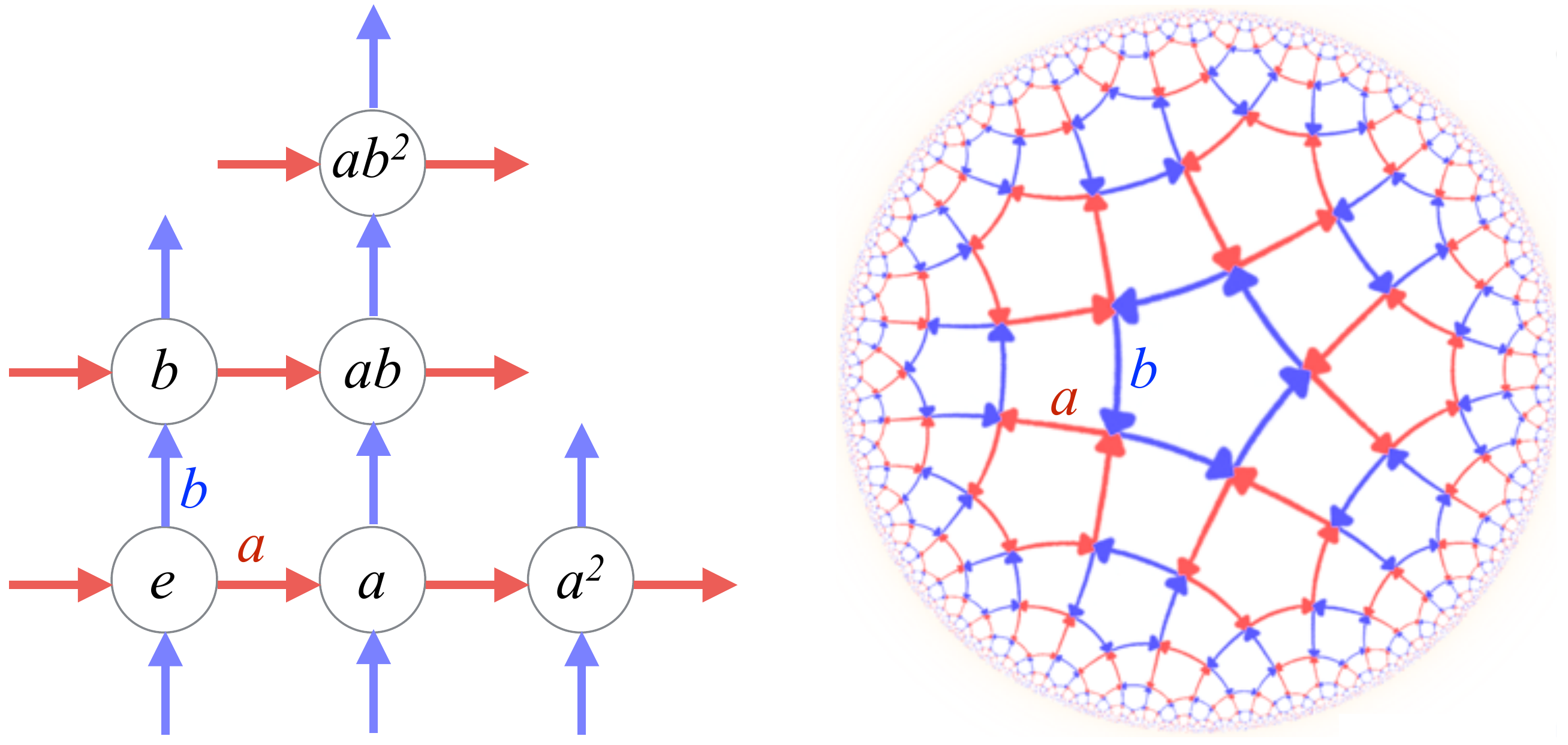}
    \caption{(colors online).  Given a group $G$ and a set $S$ of
      generators, the Cayley graph $\Gamma(G,S)$ is defined as the
      colored directed graph having set of nodes $G$, set of edges
      $\{(g,gh); g \in G, h \in S\}$, and a color assigned to each
      generator $h \in S$. {\bf Left:} the Cayley graph of the Abelian
      group $\mathbb{Z}^2$ with presentation
      $\mathbb{Z}^2=\langle a,b|aba^{-1}b^{-1}\rangle$, where a and b
        are two commuting generators. {\bf Right:} the Cayley graph of
        the non-Abelian group $G=\langle a,b|a^5,b^5,(ab)^2\rangle$.
        The Abelian-group graph is embedded into the Euclidean space
        $\mathbb{R}^2$, the non-Abelian $G$ into the Hyperbolic
        $\mathbb{H}_2$ with (negative) curvature.  }
    \label{fig:cayley}
  \end{center}
\end{figure}

The identification of an emerging spatial manifold is generally more
involved because in dimension higher than one the isometric embedding
of a discrete graph in a continuous manifold is usually impossible. However,
the notion of \emph{quasi-isometry} introduced in geometric group theory helps 
us identify the relevant geometric properties of the manifold $X$, binding the 
geometry to the algebraic properties of $X$ seen as a group. In order to clarify 
this point,  we now review the notion of quasi isometry. Given two metric spaces 
$(M_1,d_1)$ and $(M_2,d_2)$, with $d_1$ and $d_2$ the metric of the two spaces, 
a map $f:(M_1,d_1)\rightarrow (M_2,d_2)$ is a quasi-isometry if there
exist constants $A\geq 1$, $B,C\geq 0$, such that $\forall g_1,g_2\in M_1$ one has
\begin{equation*}
d_1(g_1,g_2)/A-B\leq d_2(f(g_1),f(g_2))\leq A d_1(g_1,g_2)+B,
\end{equation*}
and $\forall m\in M_2$ there exists $g\in M_1$ such that
\begin{align*}
d_2(f(g),m)\leq C.
\end{align*}
Quasi-isometry is an equivalence relation, therefore, given a Cayley graph $\Gamma$ with word
metric $d_\Gamma$, the emerging space is a manifold $(X,d_X)$ quasi-isometric to $(\Gamma,d_\Gamma)$, 
which is unique modulo quasi-isometries (see Fig. \ref{fig:cayley}). The geometric characterization of the class of metric 
spaces quasi-isometric to the Cayley graph of a group $G$ is the subject of
\emph{geometric group theory} \cite{harpe}. A crucial result is that
the quasi-isometric class is an invariant of the group, \ie\ it does
not depend on the group presentations (which instead correspond to
different Cayley graphs). Remarkably, for finitely generated groups, the quasi-isometry class 
always contains a smooth Riemaniann manifold \cite{harpe}.

A paradigmatic result \cite{gromov1984infinite} of geometric group theory is that an infinite group
$G$ is quasi-isometric to the Euclidean space $R^n$ if and only if $G$ is \emph{virtually-Abelian}, 
namely it has an Abelian subgroup $G'\subset G$ isomorphic to $\mathbb{Z}^n$ of
finite index (with a finite number of cosets).

The setting of QCAs on Cayley graphs can thus lead to a field
dynamics on either a flat spacetime or a spacetime with curvature, 
depending on whether the group $G$ is virtually Abelian or not. In the
remainder we will focus on the flat case.

\section{QCAs on Abelian groups and the small wave-vector limit}\label{s:Abelian}
In this Section we restrict to the specific subclass of automata whose
group $G$ is quasi-isometrically embeddable in the Euclidean space,
which is then virtually-Abelian. We also assume that the
representation of the isotropy group $L$ in \eqref{eq:covW} induced by
the embedding is orthogonal, which implies that the graph neighborhood
is embedded in a sphere. In words, we want homogeneity and isotropy to
hold locally also in the embedding space. Our present analysis focus
on the Abelian groups $\mathbb Z^d$ whose Cayley graphs satisfying the
isotropic embedding in the Euclidean space $\mathbb R^d$ are the
Bravais lattices. The more general scenario of virtually-Abelian
groups is discussed in Section \ref{s:future}.

In the Abelian case (and also in the virtually-Abelian case as we will
discuss in Section \ref{s:future}) it is possible to describe the
automaton in the wave-vector space. Since the group is Abelian we
label the group elements by vectors $\bg\in\mathbb{Z}^d$, and use the
additive notation for the group composition, whereas the unitary
representation of $\mathbb{Z}^d$ on $\ell_2(\mathbb{Z}^d)$ is expressed as
\begin{equation}
T_{\bh}|\bvec g\>=|\bg-\bh\>.
\end{equation}
Being the group Abelian, we can diagonalise the regular representation 
by Fourier analysis, and the operator
$A$ can be easily block-diagonalized in the wave-vector $\bk$
representation as follows
\begin{align}
  \label{eq:weylautomata} 
  A = \int_B\operatorname d^3 \! \bk  \,  |{\bk}\>\< {\bk}| \otimes
  A_{\bk},\qquad A_\bk:=\sum_{\bh\in S}e^{-i\bk\cdot\bh}A_\bh,
\end{align}
where $B$ is a compact region in $\mathbb R^3$ corresponding to the smallest 
region containing only inequivalent wave-vectors $\bk$ (usually called {\em Brillouin zone}).
Notice that the automaton is unitary if and only if $A_\bk$ unitary for every $\bk\in B$. 
The {\em plane waves} $|{\bk}\>$ on $G$ are given by
\begin{equation}
|\bk\>:=\frac{1}{\sqrt{|B|}}\sum_{\bg\in G}e^{-i\bk\cdot\bg}|\bg\>.
\end{equation}

The spectrum $\{e^{-i\omega^{(i)}_\bk}\}$ of the operator $ A_\bk$, or
more precisely its {\em dispersion relation} (namely the phases $\omega^{(i)}_\bk$ as functions of $\bk$),
plays a crucial role in the analysis of the automaton dynamics. Indeed
the speed of the wave-front of a plane wave with wave-vector $\bk$ is
given by the {\em phase-velocity} $\omega^{(i)}_\bk/|\bk|$, while the
speed of propagation of a narrow-band state having wave-vector $\bk$
peaked around the value $\bk_0$ is given by the {\em group velocity}
at $\bk_0$, namely the gradient of  $\omega^{(i)}_\bk$
evaluated at $\bk_0$.

\subsection{The small wave-vector limit}\label{ss:limit}

In order be a valid microscopic description of dynamics, the QCA model
must recover the usual phenomenology of QFT at the energy scale of
the current particle physics experiments, namely the physics of the
QCA model and the one of QFT must be the same as far as we restrict to
quantum states that cannot probe the discreteness of the underlying
lattice. For this reason it is important to address a comparison
between the automaton dynamics and the dynamics dictated by the usual
QFT differential equations. Here we show how to evaluate the behaviour
of an Abelian automaton for small wave-vectors $|\bk|\ll1$, and then
discuss a possible approach to a rigorous comparison at different
frequency scales.

The physical interpretation of the limit $|\bk|\ll1$ clearly depends on the hypotheses that we make on the order of magnitude of the QCA lattice step and time step.  As we will see in Sect.\ref{s:dirac} an heuristic argument leads us to set the scale of discreteness of the QCA at the Planck scale, thus 
the domain $|\bk|\ll1$ corresponds wavevectors much smaller than the Planck vector (consider that an ultra-high-energy cosmic ray has $k\sim10^{-8}$). Such regime corresponds to the usual one of particle physics, and is called {\em relativistic regime}.

In order to obtain the relativistic limit of an automaton $A_{\bk}$ we
define its {\em interpolating Hamiltonian} $H^A(\bk)$ as the operator
satisfying the following equality
\begin{equation}
  e^{-i H_I^A(\bk)}=A_\bk.
\end{equation}
(The term ``interpolating'' refers to the fact that the Hamiltonian $H_I^A(\bk)$ would generate a unitary evolution in continuous time that interpolates the discrete time evolution of the automaton).

Now, one can expand the Hamiltonian $H_I^A(\bk)$  to first order in $|\bk|$ 
\begin{equation}\label{eq:expansion}
  H_I^A(\bk)=H_A(\bk)+\mathcal O(|\bk|^2),
\end{equation}
corresponding to describing the evolution with the following first-order differential equation
\begin{equation}\label{eq:diff}
  i\partial_t\psi(\bk,t)=H_A(\bk)\psi(\bk,t),
\end{equation}
for narrow-band states $\psi(\bk,t)$ peaked around some $\bk_0$ with
$|\bk_0|\ll 1$.

The Hamiltonian in Eq.~\eqref{eq:diff} describes the QCA dynamics in the limit of small wave-vectors, and in the next Sections we present QCAs having the Weyl, Dirac and Maxwell Hamiltonian in as such a limit.

In Ref.~\cite{Bisio2015244} another more quantitative approach to the QFT limit of a QCA has been presented. Suppose that some automaton $A_\bk$, (with interpolating Hamiltonian $H^A_I(\bk)$) has
the unitary $U_\bk =e^{-iH_A(\bk)}$ (see Eq.~\eqref{eq:expansion}) as
first-order approximation in $\bk$. Then one can set the comparison as
a channel discrimination problem and quantify the difference between
the two unitary evolutions with the probability of error $p_e$ in the
discrimination. This probability can be computed as a function of the
discrimination experiment parameters---for example the wave-vector and
the number of particles and the duration of the evolution--- and one
can check that for values achievable in current experiments the
automaton evolution is undistinguishable from the QFT one. This
approach allows us to provide a rigorous proof that, in the limit of input states
with vanishing wave-vector, the QCA model recovers free QFT.

\section{The Weyl automaton}\label{s:weyl}

Here we present the unique QCAs on Cayley graphs of $\mathbb{Z}^d$,
$d=3,2,1$, that satisfy all the requirements of Section
\ref{s:assumptions} and with minimal internal dimension $s$ for a
non-identical evolution (see Ref.~\cite{PhysRevA.90.062106} for the
detailed derivation).

In any space dimension the only solution for $s=1$ is the identical
QCA, namely there exists no nontrivial  QCA. The minimal internal dimension for a 
non-trivial evolution is then $s=2$.

Let us start from the case of dimension $d=3$ that is the most
relevant from the physical perspective. For the group $\mathbb{Z}^3$
the only inequivalent isotropic Cayley graphs are the primitive cubic (PC) lattice, the body
centered cubic (BCC), and the rhombohedral. However only in the BCC
case, whose presentation of $\mathbb Z^3$ involves four vectors
$S_+=\{\bh_1,\bh_2,\bh_3,\bh_4\}$ with relator
$\bh_1+\bh_2+\bh_3+\bh_4=0$, one finds solutions satisfying all the
assumptions of Section \ref{s:assumptions}. There are only four
solutions, modulo unitary conjugation, that can be divided in two
pairs $A^\pm$ and $B^\pm$. A pair of solutions is connected to the
other pair by transposition in the canonical basis,
i.e. $A_\bk^\pm=(B_\bk^\pm)^T$. The first Brillouin zone $B$ for the
BCC lattice is defined in Cartesian coordinates as $-{\sqrt3}\pi\leq
k_i\pm k_j\leq{\sqrt3}\pi, \, i\neq j\in\{x,y,z\}$ and the solutions
in the wave-vector representation are
\begin{equation}\label{eq:weyl3d}
\begin{aligned}
& A^\pm_{\bk}=I u^{\pm}_\bk-i\boldsymbol\sigma^\pm\cdot
  \tilde\bn^{\pm}_\bk,\quad B_\bk^\pm=(A_\bk^\pm)^T,\\
&{\tilde\bn}^{\pm}_{\bk} :=
\begin{pmatrix}
s_x c_y c_z \mp c_x s_y s_z\\
c_x s_y c_z \pm s_x c_y s_z\\
c_x c_y s_z \mp s_x s_y c_z
\end{pmatrix},\quad
u^{\pm}_{\bk} :=  c_x c_y c_z \pm s_x s_y s_z ,\\
&  c_i:=\cos(k_i/\sqrt3),\quad s_i:=\sin(k_i/\sqrt3),
\end{aligned}
\end{equation}
where $\boldsymbol\sigma^+=\boldsymbol\sigma$ and $\boldsymbol\sigma^+=\boldsymbol\sigma^T$.

The matrices $A^\pm_\bk$ and $B^\pm_\bk$ have spectrum
$\{e^{-i\omega^{\pm}}_\bk,e^{i\omega^{\pm}}_\bk\}$ with dispersion
relation $\omega^{\pm}_\bk=\arccos(c_xc_yc_z\mp s_xs_ys_z)$ and
evolution governed by i) the wave-vector $\bk$; ii) the helicity
direction $\bvec n^{\pm}_\bk$; and iii) the group velocity $\bvec
v^\pm_\bk:=\nabla_\bk\omega^\pm_\bk$, which represents the speed of a
wave-packet peaked around the central wave-vector $\bk$.

The above solutions satisfy the isotropy constraint
and are then covariant with respect to the group
$L'$ of binary rotations around the coordinate axes, with the
representation of the group $L'$ on $\mathbb C^2$ given by
$\{I,i\sigma_x,i\sigma_y,i\sigma_z\}$. The group $L'$ is transitive on the four BCC generators of $S_+$. 

In dimension $d=2$, the only inequivalent isotropic Cayley graphs of
$\mathbb{Z}^2$ are the square lattice and the hexagonal lattice. Also
for $d=2$ we have solutions only on one of the possible Cayley
graphs, the square lattice, whose presentation of $\mathbb Z^2$
involves two vectors $S_+=\{\bh_1,\bh_2\}$. The first Brillouin zone $B$
in this case is given by $\sqrt{2}\pi\leq k_i\leq \sqrt{2}\pi,\,
i\in\{x,y\}$ and there are only two solutions modulo unitary conjugation,
\begin{equation}
\begin{aligned}\label{eq:weyl2d}
  &A_{\bk}=I u_\bk-i\boldsymbol\sigma\cdot\tilde\bn_\bk,\quad
  B_\bk:=A_\bk^T,\\
&\tilde\bn_{\bk} :=
\begin{pmatrix}
s_x c_y\\
c_x s_y\\
s_x s_y
\end{pmatrix},\quad
u_{\bk} :=  c_x c_y,\\
 & c_i:=\cos(k_i/\sqrt2),\quad s_i:=\sin(k_i/\sqrt2),
\end{aligned}
\end{equation}
with dispersion relation $\omega_\bk=\arccos(c_xc_y)$.

The QCA in Eq.~\eqref{eq:weyl2d} is covariant for the cyclic
transitive group generated by the transformation that exchanges
$\bh_1$ and $\bh_2$, with representation given by the rotation by
$\pi$ around the $x$-axis. Since the isotropy group has a reducible representation, the most general automaton is actually given by $(\cos\theta I+i\sin\theta\sigma_x)A_\bk$.


Finally for $d=1$ the unique Cayley graph satisfying our requirements
for $\mathbb Z$ is the lattice $\mathbb Z$ itself, presented as the
free Abelian group on one generator $S_+=\{h\}$. From the unitarity
conditions one gets the unique solution
\begin{align}\label{eq:weyl1d}
  A_k=u_\bk I-i\boldsymbol\sigma\cdot\tilde\bn_\bk,\qquad
\tilde\bn_{\bk} :=
\begin{pmatrix}
0\\
0\\
\sin k
\end{pmatrix},\quad
u_{\bk} :=  \cos k,
\end{align}
with dispersion relation $\omega_k =k$.

We call the solutions \eqref{eq:weyl3d}, \eqref{eq:weyl2d} and
\eqref{eq:weyl1d} Weyl automata, because in the limit of small
wave-vectors of Section \ref{ss:limit} their evolution obeys
Weyl's equation in space dimension $d=3$, $d=2$ and $d=1$,
respectively. All the previous solution in Eqs. \eqref{eq:weyl3d}, \eqref{eq:weyl2d}, and \eqref{eq:weyl1d} for dimension $d=3,2,1$ can be rewritten in the form 
\begin{align}
W_\bk=u_\bk I-i\boldsymbol\sigma\cdot\tilde\bn_\bk,
\end{align}
for certain $u_k$ and $\bn_k$, with dispersion relation
\begin{align}
\omega_\bk=\arccos{u_\bk}.
\end{align}\label{eq:weyl-interpolating}
It is easily to check that the interpolating Hamiltonian is
\begin{align}
H^W_I(\bk)=\boldsymbol\sigma\cdot\bn_\bk,\qquad \bn_\bk:=
\frac{\omega_{\bk}}{\sin\omega_{\bk}}\tilde{\bn}_{\bk},
\end{align}
and by power expanding at the first order in $\bk$ one has
\begin{align}
  H_I^{W}=H_{W}(\bk)+\mathcal O(|\bk|^2), \qquad
  H_{W}(\bk)=\tfrac{1}{\sqrt{d}} \boldsymbol\sigma\cdot \bk
\end{align}
where $H_{W}(\bk)$ coincides with the usual Weyl Hamiltonian in
$d$ dimensions once the wave-vector $\bk$ is interpreted as the momentum.

\section{The Dirac automaton}\label{s:dirac}

From the previous section we know that in our framework all the admissible QCAs with $s=2$ gives the Weyl equation in the limit of
small wave-vectors. In order to get a more general dynamics---say the Dirac one---it is then necessary to increase the internal degree of
freedom $s$. Instead of deriving the most general QCAs with $s>2$, in Ref.~\cite{PhysRevA.90.062106} it is shown how the Dirac equation for any space dimension $d=1,2,3$ can be derived from the local coupling of two Weyl automata. Here we shortly review this result.

Starting from two arbitrary Weyl automata $W$ and $W^\prime$ in
dimension $d$ (see the solutions \eqref{eq:weyl3d}, \eqref{eq:weyl2d}
and \eqref{eq:weyl1d} in for $d=3,2,1$, respectively),
the coupling is obtained by performing the direct-sum of their
representatives $W_{\bk}$ and ${W'_\bk}$, obtaining a QCA with $s=4$,
and introducing off-diagonal blocks $X$ and $Y$ in such a way that the
obtained matrix is unitary. The locality of the coupling implies that
the off-diagonal blocks are independent of $\bk$, namely
\begin{equation}
  D_\bk:=
  \begin{pmatrix}
    p{W_\bk}&qX\\
    rY&t{W'_\bk}
  \end{pmatrix},\qquad p,q,r,t\in\mathbb{C}.
  \label{eq:diracstart}
\end{equation}

In order to satisfy all the hypothesis of Section \ref{s:assumptions}
it is possible to show that the unique local coupling of Weyl QCAs,
modulo unitary conjugation, are
\begin{equation}\label{eq:dirac-gen}
  D_\bk:=
  \begin{pmatrix}
    n W_\bk&im\\
    im&nW_\bk
  \end{pmatrix},\qquad n,m\in\Reals^+,\quad n^2+m^2=1,
\end{equation}
which are conveniently expressed in terms of gamma matrices in the
spinorial representation as follows
\begin{align}\label{eq:dirac}
  D_\bk:= nI u_\bk-i n\gamma^0\boldsymbol\gamma\cdot\tilde\bn_\bk+im\gamma^0,
\end{align}
where the functions $u_\bk$ and $\tilde\bn_\bk$ depend on the
value of $d$ for the Weyl automaton $W_\bk$ in Eq.~\eqref{eq:dirac-gen}.
Notice the dispersion relation of the QCAs \eqref{eq:dirac} that is
simply given by
\begin{equation}
  \omega_\bk=\arccos[\sqrt{1-m^2}u_\bk].
\end{equation}

The QCAs in Eq.~\eqref{eq:dirac-gen} in the small wave-vector limit and for $m\ll 1$ all give the usual Dirac equation in the respective dimension $d$, with $m$ corresponding to the particle mass. Indeed, the interpolating Hamiltonian $H_I^{D}(\bk)$ is given by
\begin{align}
  H_I^{D}(\bk)=f(\bk)(n\gamma^0\boldsymbol\gamma\cdot\tilde\bn_\bk-m\gamma^0)
  ,\; f(\bk):=
  \frac{\omega_{\bk}}{\sin\omega_{\bk}},
\end{align}
that by power expanding at the first order in $\bk$ is approximated as follows
\begin{align}
  &H_I^{D}(\bk)=H_{D}(\bk)+\mathcal O(|\bk|^2), \nonumber\\
  &H_{D}(\bk)=\frac n{\sqrt d}\gamma^0\boldsymbol\gamma \cdot\bk+m\gamma^0.
\end{align}
Finally, for small values of $m$, $m\ll1$, we have $n\simeq 1+\mathcal
O(m^2)$ and neglecting terms of order $\mathcal O(m^2)$ and $\mathcal
O(|\bk|^2)$
\begin{equation}
  H_I^{D}(\bk)=\frac 1{\sqrt d}\gamma^0\boldsymbol\gamma \cdot\bk+m\gamma^0+O(m^2)+O(|\bk|^2),
\end{equation}
one has the Dirac Hamiltonian with the wave-vector $\bk$ and the
parameter $m$ interpreted as momentum and mass, respectively.
For $d=1$, modulo a permutation of the
canonical basis, the QCA corresponds to two identical and
decoupled $s=2$ automata. Each of these QCAs coincide with the one
dimensional Dirac automaton derived in Ref.~\cite{Bisio2015244}. The
last one was derived as the simplest ($s=2$) homogeneous QCA covariant
with respect to the parity and the time-reversal transformation, which
are less restrictive than isotropy that singles out the only Weyl QCA
\eqref{eq:weyl1d} in one space dimension.

We want to emphasize that in the above derivation everything is adimensional by construction. Dimensions can be recovered by providing  values $\tau$ and $a$ in seconds and meters, respectively, to the discreteness scales in time and space of the QCA, and providing the maximum value of the mass $M$ in kilograms  corresponding to $|m|=1$ in Eq. \eqref{eq:dirac-gen}. From the relativistic limit, the comparison with the usual dimensional Dirac equation leads to the identities $c=a/\tau$, $\hbar=Mac$, which leave only one unknown among the three variables $a,\tau,$ and $M$. At the maximum value of the mass $|m|=1$ in Eq. \eqref{eq:dirac-gen}  we get a non evolving automaton, with a flat dispersion relation, which can be interpreted as a mini black-hole, where the Schwarzild radius equals the localization length, i.e. the Compton wavelength, corresponding to a mass equal to the Planck mass. We thus heuristically interpret $M$ as the Planck mass, and from the two identities  $c=a/\tau$, $\hbar=Mac$ we get the Planck scale.

\section{QCA for free electrodynamics}\label{s:maxwell}

In Sections \ref{s:weyl} and \ref{s:dirac} we showed how the
dynamics of free Fermionic fields can be derived within the QCA
framework starting from informational principles. Within this
perspective the information contained in a finite number of systems
must be finite and this is the reason why we consider Fermionic QCAs. 
One might then wonder how the physics of the free electromagnetic 
field can be recovered in this framework, and more generally any
Bosonic quantum field obeying the canonical commutation relations.
In the present section we review the results of Ref.~\cite{bisio2014quantum},
where the above question was answered in detail.

The basic idea behind this approach is to model the photon as an
entangled pair of Fermions evolving according to the Weyl QCA
presented in Section \ref{s:weyl}. Then we show that in a suitable
regime both the free Maxwell equation in three dimensions and the
Bosonic commutation relations are recovered. For this purpose, we 
consider two Fermionic fields, which in the wave-vector
representation are denoted as $\psi(\bk)$ and $\varphi(\bk)$. The
evolutions of these two fields are given by
\begin{align}
  \label{eq:automa2}
  {\psi} (\bk,t+1) = W_\bk{\psi} (\bk,t).
\quad
  {\varphi} (\bk,t+1) = W_\bk^*{\varphi} (\bk,t).
\end{align}
Where the matrix $W_\bk$ can be any of the Weyl QCAs in three space
dimensions of Eq.~\eqref{eq:weyl3d}, (the whole derivation is
independent on this choice) and $W_\bk^* = \sigma_y
  W_\bk\sigma_y$ denotes the complex
conjugate matrix.

We now introduce the following bilinear operators
\begin{align}
  \label{eq:prephoton}
 \bvec{G}_T(\bk,t) &:=  \bvec{G}(\bk,t) -  
\left(\frac{\bvec{n}_{\frac{\bk}{2}}}{|\bvec{n}_{\frac{\bk}{2}}|} \cdot
{\bvec{G}}(\bk,t) \right)
\frac{\bvec{n}_{\frac{\bk}{2}}}{|\bvec{n}_{\frac{\bk}{2}}|} \\
\bvec{G}(\bk,t) &:= ( G^{1}(\bk,t), G^{2}(\bk,t), G^{3}(\bk,t))^T \nonumber\\
  G^{i}(\bk,t) &:= \varphi^T (\tfrac{\bk}{2},t) \sigma^{i}  \psi(\tfrac{\bk}{2} , t)\nonumber\\
  &=\varphi^T (\bk,0) ( W_{\tfrac{\bk}{2}}^\dag  \sigma^{i}
  W_{\tfrac{\bk}{2}} )\psi(\tfrac{\bk}{2} , 0) 
\nonumber
\end{align}
with $\bn_\bk$ as in Eq.~\eqref{eq:weyl-interpolating}. By
construction the field $\bvec{G}_T(\bk,t)$ satisfies the following relations
\begin{align}
  \label{eq:premaxwell}
 \bvec{n}_{\frac{\bk}{2}} \cdot \bvec{G}_T(\bk,t)  &=  0,\\
  \bvec{G}_T(\bk,t) &= \Exp(-i2\bvec{n}_{\tfrac{\bk}{2}} \cdot \bvec{J}
  t)  \bvec{G}_T(\bk,0), \label{eq:premaxwell2}
\end{align}
where we used the identity 
$ \exp (-\tfrac{i}{2}\bvec{v}\cdot \boldsymbol{\sigma}) 
\boldsymbol{\sigma} \exp (\tfrac{i}{2}\bvec{v} \cdot \boldsymbol{\sigma}) =
\Exp(-i\bvec{v} \cdot \bvec{J}) \boldsymbol{\sigma} $,
the matrix $\Exp(-i\bvec{v}\cdot\bvec{J})$ acting on $\boldsymbol{\sigma}$ regarded as a vector,
and $\bvec J=(J_x, J_y,J_z)$ representing the infinitesimal generators of $\SU(2)$ in the spin 1 representation.
Taking the time derivative of Eq. \eqref{eq:premaxwell2} we obtain
\begin{align}
  \label{eq:premaxwell3}
  \partial_t\bvec{G}_T(\bk,t) = 2\bvec{n}_{\tfrac{\bk}{2}} \times  \bvec{G}_T(\bk,t).
\end{align}
If $\bvec{E}_G$ and
$\bvec{B}_G$ are two Hermitian operators defined by the relation
\begin{align}
  \label{eq:electric and magnetic field}
  \bvec{E}_G:=|{\bn}_{\tfrac\bk2}|(\bvec{G}_T+\bvec{G}_T^\dag),\quad\bvec{B}_G:=i|{\bn}_{\tfrac\bk2}|(\bvec{G}_T^\dag-\bvec{G}_T),
\end{align}
then Eq. \eqref{eq:premaxwell} and Eq. \eqref{eq:premaxwell3}  
can be rewritten as 
\begin{align}
   & \partial_t \bvec{E}_G = i 2\bvec{n}_{\tfrac{\bk}{2}} \times
    \bvec{B}_T(\bk,t)  
&&\partial_t \bvec{B}_G =- i 2\bvec{n}_{\tfrac{\bk}{2}} \times \bvec{E}_T(\bk,t)    \nonumber\\
          & 2\bvec{n}_{\tfrac{\bk}{2}} \cdot \bvec{E}_G = 0 
&& 2\bvec{n}_{\tfrac{\bk}{2}} \cdot \bvec{B}_G = 0 
      \label{eq:maxweldistorted}
 \end{align}
 that are the free Maxwell's equation in the wave-vector space with the
 substitution $ 2\bvec{n}_{\tfrac{\bk}{2}} \to \bk$.  In the limit
 $|\bk| \ll 1$ one has $ 2\bvec{n}_{\tfrac{\bk}{2}} \sim \bk$ and the
 usual free electrodynamics is recovered.

 However the field defined in Eqs. \eqref{eq:prephoton} and
 \eqref{eq:electric and magnetic field} does not satisfy
 the correct Bosonic commutation relations.  As shown in
 Ref.~\cite{bisio2014quantum} the solution to this problem is to
 replace the operators $G^i$ defined in Eq. \eqref{eq:prephoton} with
 the operators $F^i$ defined as
\begin{align}
  \label{eq:photon}
F^{i}(\bk) := 
 \int \frac{ d \bvec{q}}{(2 \pi)^3}
f_{\bk}(\bvec{q})
\varphi
\left(\tfrac{\bk}{2}-\bvec{q}\right)
\sigma^{i}
\psi
 \left(\tfrac{\bk}{2}+\bvec{q}\right) 
\end{align}
where
$\int\frac{d\bvec{q}}{(2\pi)^3} |f_{\bk}(\bvec{q})|^2 =1, \forall \bk$.
In terms of $\bvec{F}(\bk)$, we can define the polarization operators 
$\gamma^i(\bk)$
of the electromagnetic field as follows
\begin{align}
 &\gamma^i(\bk) := \bvec{u}^i_\bk\cdot\bvec{F}(\bk,0),\quad i=1,2,
  \label{eq:polarization}
\\
&\bvec{u}^i_\bk \cdot \bn_{\bk} =\bvec u^1_\bk\cdot\bvec u^2_\bk= 0,
\;
 |\bvec u^i_\bk|=1,
\;
(\bvec u^1_\bk\times\bvec u^2_\bk)\cdot\bn_\bk>0.
\end{align}
In order to avoid the technicalities of the continuum we suppose to have a discrete wave-vector space (as if the electromagnetic field were confined in a finite volume) and moreover let us assume $|
{f}_\bk(\bvec{q})|^2$ to be a constant function over a region
$\Omega_\bk$ which contains $N_\bk$ modes, i.e. $|{f}_\bk(\bvec{q})|^2 =\tfrac{1}{N_\bk}$ if $\bvec{q}\in \Omega_\bk$ and $|{f}_\bk(\bvec{q})|^2 = 0 $ if $\bvec{q} \not\in \Omega_\bk$.  Then, for a given state $\rho$ of the field we denote by $M_{\varphi,\bvec{k}}$ (resp.  $M_{\psi,\bvec{k}}$) the mean number of type $\varphi$ (resp $\psi$) Fermionic excitations in the region $\Omega_\bk$.  One can then show that, for states such that $M_{\xi,\bvec{k}}/ N_\bk \leq
\varepsilon$ for all $\xi = \varphi, \psi$ and $\bvec{k}$ and for $\varepsilon \ll 1$ we can safely assume $ [\gamma^i
(\bvec{k}),{\gamma^j}^\dag (\bvec{k}')]_- = \delta_{i,j}
\delta_{\bvec{k},\bvec{k}'}$, i.e. the polarization operators are Bosonic operators.

\section{Future perspectives: interacting QCAs and gravity}\label{s:future}

In the previous sections we showed how the dynamics of free relativistic quantum fields
emerges from the evolution of states of Fermionic QCAs, provided that they satisfy the
requirements of unitarity, linearity, homogeneity and isotropy. However, in order to recover
relativistic quantum field theory we need to find also interacting evolutions, where 
Fermions and Bosons can scatter, with Bosonic fields carrying the fundamental interactions.
For this purpose, one needs to overcome the linearity assumption, allowing Fermionic 
excitations to exchange momentum. There is a very good reason to introduce a non-linear 
evolution, which is precisely due to the discrete nature of the QCA evolution. 
Indeed, while in a context where time is continuous it makes sense to require that the canonical 
basis in the Hilbert space representing a local system \footnote{In the present case local systems are local Fermionic modes. 
For a detailed description of the Fermionic theory, see 
Refs.~\cite{0295-5075-107-2-20009,doi:10.1142/S0217751X14300257}.} changes 
continuously in time, when the evolution occurs in discrete steps, as in a QCA, there is no natural way 
to compare the local reference system at  subsequent times, and it is thus necessary to allow for an 
uncontrollable misalignment of the local reference frame. One can then introduce a completely local 
non-linear evolution at each step, following the linear one, preserving homogeneity, 
isotropy and unitarity. This misalignment provides a natural notion of a 
quantum gauge symmetry, with free evolution of the gauge field dictated by the structure of the local unitary
QCA. In this way, one does not need to artificially quantize the gauge fields, 
nor introduce the free Bosonic Hamiltonian or Lagrangian. This generalization is expected to provide 
an effective description corresponding to different fundamental interactions,
possibly including a fully quantum spontaneous symmetry breaking mechanism providing mass to the massless Fermions. 
In this case, we would have a dynamical mechanism instead of the construction that we showed in Sect.~\ref{s:dirac}, 
which would then be an effective representation. The study of this mechanism along with its symmetries is also 
expected to provide a reasonable attempt at the formulation of a quantum theory of gravity, as it relates the
symmetries of the mechanism lying at the core of mass with the symmetries of the emergent space-time,
suggesting a relation between geometry and interactions of quantum fields.

\begin{acknowledgements}
The authors acknowledge stimulating and fruitful discussion with R. Sorkin. This work has been supported in part by the 
Templeton Foundation under the project ID\# 43796 {\em A Quantum-Digital Universe}.
\end{acknowledgements}


\end{document}